\documentclass[sigconf]{acmart}

\newcommand{\cpp}{{C\nolinebreak[4]\hspace{-.05em}\raisebox{.4ex}{\small\bf ++ }}}%this version seems to look better
\usepackage{tabularx}
\usepackage{siunitx} 
\usepackage{cleveref}
\usepackage{enumitem}
\usepackage{framed}
\usepackage{listings}
\usepackage{multirow}
\usepackage{subcaption}

\newcommand{\enquote}[1]{``#1''}

\AtBeginDocument{%
  }

\copyrightyear{2026}
\acmYear{2026}
\setcopyright{cc}
\setcctype{by}
\acmConference[FTW '26]{3rd International Flaky Tests Workshop 2026 }{April 12--18, 2026}{Rio de Janeiro, Brazil}
\acmBooktitle{3rd International Flaky Tests Workshop 2026 (FTW '26), April 12--18, 2026, Rio de Janeiro, Brazil}
\acmDOI{10.1145/3786178.3788327}
\acmISBN{979-8-4007-2409-1/2026/04}

\begin{document}

%\title{An LLM-Based Analysis of Issue Reports Related to Fixed Flaky Tests in a Large Industrial Database Management System}
\title{Flaky Tests in a Large Industrial Database Management System: An Empirical Study of Fixed Issue Reports for SAP HANA}

\author{Alexander Berndt}
\email{alexander.berndt@uni-heidelberg.de}
\orcid{0009-0009-5248-6405}
\affiliation{%
  \institution{Heidelberg University}
  \country{Germany}
}
\additionalaffiliation{SAP}

\author{Thomas Bach}
\email{thomas.bach03@sap.com}
\orcid{0000-0002-9993-2814}
\affiliation{%
  \institution{SAP}
  \country{Germany}
}

\author{Sebastian Baltes}
\email{sebastian.baltes@uni-heidelberg.de}
\orcid{0000-0002-2442-7522}
\affiliation{%
  \institution{Heidelberg University}
  \country{Germany}
}
% \email{alexander.berndt@uni-heidelberg.de}
% \orcid{0009-0009-5248-6405}
% \affiliation{%
%   \institution{Heidelberg University}
%   \country{Germany}
% }

% \author{Thomas Bach}
% \email{thomas.bach03@sap.com}
% \orcid{0000-0002-9993-2814}
% \affiliation{%
%   \institution{SAP}
%   \city{Walldorf}
%   \country{Germany}
% }

% \author{Sebastian Baltes}
% \email{sebastian.baltes@uni-heidelberg.de}
% \orcid{0000-0002-2442-7522}
% \affiliation{%
%   \institution{Heidelberg University}
%   \country{Germany}

\begin{abstract}
  Flaky tests yield different results when executed multiple times for the same version of the source code. Thus, they provide an ambiguous signal about the quality of the code and interfere with the automated assessment of code changes. While a variety of factors can cause test flakiness, approaches to fix flaky tests are typically tailored to address specific causes. However, the prevalent root causes of flaky tests can vary depending on the programming language, application domain, or size of the software project. Since manually labeling flaky tests is time-consuming and tedious, this work proposes an LLMs-as-annotators approach that leverages intra- and inter-model consistency to label issue reports related to fixed flakiness issues with the relevant root cause category. This allows us to gain an overview of prevalent flakiness categories in the issue reports.
  We evaluated our labeling approach in the context of SAP HANA, a large industrial database management system. 
  Our results suggest that SAP HANA's tests most commonly suffer from issues related to concurrency (23\%, 130 of 559 analyzed issue reports). Moreover, our results suggest that different test types face different flakiness challenges. Therefore, we encourage future research on flakiness mitigation to consider evaluating the generalizability of proposed approaches across different test types.
\end{abstract}





\maketitle

\section{Introduction}
\label{sec:introduction}

Flaky tests yield different results when executed multiple times for the same version of the source code. % under stable environmental conditions.
Thus, flaky tests affect various aspects of test automation and represent a major problem for industrial software testing~\cite{berndt2024taming,gruber2022survey,harman2018startups,hoang2024presubmit}. They hinder automatic merging of code changes, reduce the effectiveness of test prioritization techniques, and erode developers' trust in test results~\cite{fallahzadeh2022impact,gruber2022survey,berndt2024taming}. 

Previous research has studied the root causes of flakiness and proposed strategies to detect and fix flaky tests~\cite{eck2019understanding,luo2014empirical,parry2022survey,fatima2024flakyfix,liu2024WEFix,berndt2023vocabulary,berndt2024test,magill2025deflake,haben2024importance}. Due to the range of potential issues that can cause flakiness, there seems to be no one-size-fits-all solution that can be applied to detect or fix all types of flakiness. Therefore, specialized solutions have emerged that are often tailored to specific root causes of flakiness~\cite{rahman2024flakesync,zhang2021domainspecific,chen2024neurosymbolic,liu2024WEFix}. Hence, before assessing existing approaches to fix test flakiness, it is beneficial to know the root causes of flaky tests. Although previous studies identified prevalent root causes of flaky failures~\cite{schroeder2025preliminary,luo2014empirical,romano2021empirical,parry2022survey}, prior research also suggests that the root causes of flakiness vary between different programming languages and types of software~\cite{gruber2022survey,schroeder2025preliminary}. Therefore, results from previous studies cannot always be generalized to very large software projects that pose unique challenges~\cite{bach2022testing,memon2017taming}. 

Therefore, we performed an empirical analysis of flakiness-related issue reports to gain an overview of the root causes of flaky tests in the context of SAP HANA. SAP HANA is a large industrial database management system that serves as the foundation for SAP's business applications~\cite{may2025saphanacloud}. The SAP HANA test suite consists primarily of two test types: native \cpp unit tests and Python-based system tests that interact with a running database instance via SQL.
To gain insights into differences in flakiness between test types, we categorized issue reports by affected test type using large language models (LLMs) as annotators. Based on the labeled issue reports, we identified \textsc{Concurrency} as the most prevalent root cause of flakiness in the given dataset, appearing in 130 of 559 issue reports (23\%). 
Based on a comparison with a manually labeled sample of $n=50$ issue reports, our results indicate that LLMs can enable large-scale flakiness analyses in industrial contexts. Our investigation of cases where manual labels do not match those generated by LLMs shows that some flakiness categories can overlap within a single test. This finding leads to the conclusion that identifying root causes of flaky tests is a multi-label task.  

% Comparing system tests and unit tests, we find that system tests commonly suffer from \emph{Oracle Brittleness} due to missing \texttt{ORDER BY} or brittle string matching. In contrast, unit tests tend to be more susceptible to \emph{Platform} differences such as compiler differences. 

% Based on an analysis of issue reports over time, we find that \emph{Concurrency} consistently leads to a high rate of related issue reports, causing 10 or more related issue reports per quarter in 9 out of 11 quarters. We observe the highest increase in flakiness-rated issue reports for \emph{Async wait}, where a single quarter contains 13 issue reports, while the other quarters range from 1 to 7. 

In summary, this work provides the following contributions:
\begin{enumerate}
    \item The evaluation of an LLM-as-annotators approach for a large-scale flakiness analysis.
    \item A comparison of the root causes of flakiness between different test types in a very large software project.
    \item An overview of how the numbers of flakiness issue reports evolve over time in a large industrial environment.
\end{enumerate}

The remainder of this document is structured as follows.
We start by introducing the relevant context in \Cref{sec:related-work} and \Cref{sec:background}, followed by a description of our dataset in \Cref{sec:dataset}. In \Cref{sec:methodology}, we motivate our research questions and the respective methodology before presenting our results in \Cref{sec:results}. Finally, we discuss our findings in \Cref{sec:discussion} and conclude this work in \Cref{sec:conclusion}.

\section{Related Work}
\label{sec:related-work}

In \Cref{tab:flakiness-categories}, we describe the flakiness categories we used to categorize the issue reports in this work. In the following, we introduce related work on categorizing flaky tests. Note that we adapted the naming of the root cause categories to the language used by SAP developers, as described in the following.

\begin{table}
\centering
\small
\caption{Flakiness categories derived from related work.}
\label{tab:flakiness-categories}
\begin{tabularx}{\linewidth}{p{1.75cm}X}
\toprule
\textbf{Category} & \textbf{Description} \\
\midrule
\textsc{Concurrency} & Test failures due to timing issues, thread handling problems, deadlocks, or synchronization issues in multi-threaded or distributed environments~\cite{luo2014empirical}. \\
\textsc{Timeout} & Tests that exceed time limits, e.g., due to system load or slow hardware, causing flaky failures~\cite{berndt2024taming}. \\
\textsc{Oracle-\newline Brittleness} & Assumptions about ordering, exact error messages, or non-deterministic checks; lack of robustness against minor behavioral differences~\cite{parry2022what}. \\
\textsc{Configuration} & Incorrect or missing configuration parameters that are not aligned with test expectations~\cite{presler2019wait}. \\
\textsc{Async Wait} & Test failures due to asynchronous waits for external resources, which are not handled robustly~\cite{luo2014empirical}. \\
\textsc{Isolation} & Inter-test dependencies, e.g., where tests lack proper cleanup and interfere with other tests, causing cascading failures~\cite{parry2022survey}. \\
\textsc{Platform} & Differences related to processor architecture, compiler behavior, or alignment issues~\cite{eck2019understanding}. \\
\textsc{Application} & Failures due to defects in the tested application~\cite{googletestingblog}. \\
\textsc{Test-\newline Framework} & Issues with testing infrastructure, including test harnesses, assertion frameworks, or utilities~\cite{presler2019wait,googletestingblog,osikowicz2025empirically}. \\
\textsc{Error-\newline Handling} & Missing or improper handling of exceptions leading to unexpected behavior~\cite{fatima2024flakyfix}. \\
\textsc{Memory-\newline Management} & Failures related to improper memory handling, e.g., use of uninitialized variables or use-after-free~\cite{gruber2022survey,chromium}. \\
\textsc{Environment} & Test failures, e.g., due to slow or constrained filesystems, that cause flaky failures~\cite{vahabzadeh2015empirical}. \\
\textsc{Randomization} & Tests relying on random behavior without deterministic sampling, leading to flaky outcomes~\cite{eck2019understanding}. \\
\textsc{Network} & Transient network errors, connection resets, or flaky inter-service communication~\cite{luo2014empirical}. \\
\textsc{Numeric-\newline Semantics} & Numeric inaccuracies caused by limited floating-point precision causing flaky failures~\cite{eck2019understanding}. \\
\textsc{Locale} & Timezone differences affecting test expectations~\cite{eck2019understanding}. \\
\bottomrule
\end{tabularx}
\end{table}

\citet{luo2014empirical} conducted an analysis of flaky tests using 201 flakiness-fixing commits from 51 open-source repositories, mostly written in Java. Based on the examined commits, they identified \textsc{Async wait}, \textsc{Concurrency}, and `test order dependency' (\textsc{Isolation} in our taxonomy) as the most common root causes for flakiness. \citet{eck2019understanding} also found that \textsc{Async wait} and \textsc{Concurrency} are the most common root causes based on an analysis of 200 flaky tests reported in Mozilla's issue tracker. In addition to \citet{luo2014empirical}, they identified `too restrictive range' (\textsc{Oracle Brittleness}) as the third common category, appearing in 40 of 234 cases (17\%). 

\citet{romano2021empirical} adopted the methodology introduced by \citet{luo2014empirical} to gain insights into the flakiness of user interface (UI) tests. Based on an analysis of commit messages, bug reports, and code changes for 235 flaky UI tests in 62 open-source projects, they found \textsc{Async wait} to be the most prominent category. However, in contrast to \citet{luo2014empirical}, issues related to the execution environment and the test runner for UI tests were the second and third most common root causes of flakiness. This difference suggests that the root causes of flakiness vary between test targets, which developers also reported in existing surveys~\cite{gruber2022survey,eck2019understanding}. 
Furthermore, results from a developer survey suggested that software for different domains may also suffer from distinct root causes of flakiness~\cite{gruber2022survey}.
Studies on flakiness in Python or Rust provided additional evidence that some root causes of flakiness may be specific to certain programming languages~\cite{schroeder2025preliminary,gruber2021empirical}. For example, categories such as \textsc{Uninitialized variables} can be particularly relevant for languages with lower-level memory management, such as \cpp~\cite{gruber2022survey}.

\begin{figure}
  \centering
  \fbox{
    \begin{minipage}{\linewidth} % Adjust the width as needed
      \textbf{Description:} \texttt{<test\_suite\_j>} fails FAIL: \texttt{<test\_x>} \\
      \textbf{Root Cause:} Ordered results expected, but unordered results produced. \\
      \textbf{Symptom:} \texttt{<test\_x>} fails sporadically. \\
      \textbf{Resolution:} Expect unordered results.
    \end{minipage}
  }
  \caption{An example issue report.}
  \label{fig:report-example}
\end{figure}

\section{Study Context}
\label{sec:background}
SAP HANA is a large-scale in-memory database management system that has been developed for more than 10 years, consisting of millions of lines of code~\cite{bach2022testing}. The SAP HANA repository is maintained by more than \num{100} active developers.~\cite{bach2022testing}.
Since SAP HANA commonly serves as the data management backbone of SAP's enterprise software~\cite{may2025saphanacloud}, potential software failures can incur high costs \cite{herzig2015art}. To prevent such failures, SAP HANA is continuously and extensively tested. 

SAP HANA's test suite consists of two types of tests: \emph{native unit tests}, written in \cpp, and Python \emph{system tests} that communicate with a running database instance via SQL. 
Typically, unit tests have \enquote{low} execution times of a few seconds, whereas system tests exhibit \enquote{medium to large} execution times up to multiple hours~\cite{bach2019dynamic}. Developers consider system tests to be more heavily affected by flakiness due to their larger scope and lower degree of isolation~\cite{berndt2024taming,parry2022survey}. Likewise, in SAP HANA's CI system, test failures  trigger three repeated executions of the failing test to prevent breakages caused by flaky failures. Thus, flaky failures of system tests are also considered more costly, as they lead to higher wait times for developers and require more computational resources due to their increased execution times. In contrast to the typical testing pyramid, where a higher number of unit tests builds the foundation of a test suite, SAP HANA's developers focus more on system tests, as they provide valuable information on the system under test. Thus, both types of test play an important role in testing SAP HANA.  

% Potential issues with SAP HANA are typically reported in an issue tracking system. The resulting tickets contain information such as a priority, tags, keywords, and related components. We utilize the fields \emph{Description}, \emph{Root cause}, \emph{Symptom}, and \emph{Resolution} for our analysis. The \emph{Description} field typically contains a sentence describing the issue. The \emph{Root cause} contains a description of the reason for the issue. \emph{Symptom} describes the unexpected behavior resulting from the issue. If the issue has been fixed, the developers record the fix in \emph{Resolution}.

\section{Dataset}
\label{sec:dataset}

\begin{figure}
    \centering
\includegraphics[width=\linewidth]{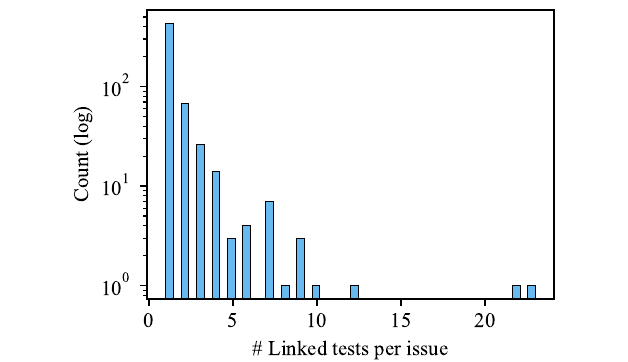}
    \caption{Histogram showing the number of linked tests per issue report. Note the logarithmic scale of the y-axis.}
    \label{fig:tests-per-report}
\end{figure}

We selected a dataset of 559 issue reports of fixed flaky tests sampled over a 34-month period between 2023 and 2025.
These issue reports were tagged as related to flakiness and marked \enquote{resolved fixed} in the issue tracker. Each of the issue reports was linked to one or more tests. \Cref{fig:tests-per-report} shows the distribution of test count per report. The number of tests per issue report ranges from 1 to 23. The issue reports were most commonly related to a single test (434 of 559 issue reports, 78\%). Since some tests appear in multiple reports, our study covers 587 unique tests, each labeled either as a \emph{system test} or a \emph{native unit test}.
We assigned each issue report to the respective test type (\emph{system test}, \emph{native unit test}, or \emph{mixed}), yielding 439 issues related to system tests, 108 to native unit tests, and 12 mixed. 

We utilized the following issue fields for this study: \emph{Description}, \emph{Root cause}, \emph{Symptom}, and \emph{Resolution} as shown in \Cref{fig:report-example}. The fields for root cause and description are filled out when an issue report is filed and were filled out in all issue reports. In contrast, symptoms and resolutions are entered after an issue has been fixed. The symptom of the issue was available in 95\% of the reports and the resolution was available in 73\% of the reports.

\section{Research Questions and Methodology}
\label{sec:methodology}

In this section, we introduce the research questions and the methodology we used to answer them.

\begin{description}[style=multiline, labelindent=4mm, leftmargin=12mm, topsep=4pt]
\item[RQ1] What are the prevalent root causes of flaky tests in SAP HANA issue reports for unit vs. system tests?
\item[RQ2] Can we identify trends or patterns in the root causes of flaky tests in SAP HANA issue reports?
    \begin{description}[style=multiline, leftmargin=10mm]
    \item[RQ2.1] How do root causes of flaky tests differ in issue reports linked to multiple tests? % in the context of SAP HANA?
    \item[RQ2.2] How do root causes of flaky tests change over time? % in the context of SAP HANA?
    \end{description}
\end{description}

\subsection{RQ1: Root Causes}

\subsubsection{Motivation}

As described in \Cref{sec:background}, previous research indicates that the most common root causes of flakiness can vary depending on the type of test~\cite {gruber2022survey}.
Furthermore, most of the existing approaches to automatically fix flaky tests are customized to fix specific root causes of flakiness~\cite{rahman2024flakesync,zhang2021domainspecific,chen2024neurosymbolic,liu2024WEFix}; there seems to be no one-size-fits-all solution readily available.
Therefore, before evaluating potential approaches to fix flakiness in the context of SAP HANA, our objective was to gain an overview of the root causes of flakiness. As described in \Cref{sec:background}, both system and native unit tests play an important role in testing SAP HANA. With the first research question, our aim is to determine whether there are differences in the root causes of flakiness between the two test types.

\subsubsection{Approach}

To gain an overview of flakiness root causes, we analyzed issue reports in SAP HANA's ticket system that are related to flaky tests. 
Similar to previous work~\cite{schroeder2025preliminary}, we focused our analysis on issue reports where the root cause of flakiness is already known, i.e., issue reports with the status \enquote{resolved fixed}. To analyze issue reports, our approach combined automated and manual methods. First, we have manually annotated a subset of 232 tests to establish a ground truth for estimating the success rate of our automated approach. This sample size represents the minimum required to ensure that the resulting estimate lies within a 95\% confidence interval (5\% margin of error). Second, given an issue report, we automatically labeled the affected test(s) linked to the report as system or native unit tests based on the content of the respective test file. Comparing with the ground truth that we manually annotated, we obtained an accuracy of 96\%, which we considered sufficient to proceed with our analysis. 
Finally, our automated approach clustered the 587 tests into 464 system tests and 123 \cpp tests.

\begin{figure}
    \centering
    \includegraphics[width=\linewidth]{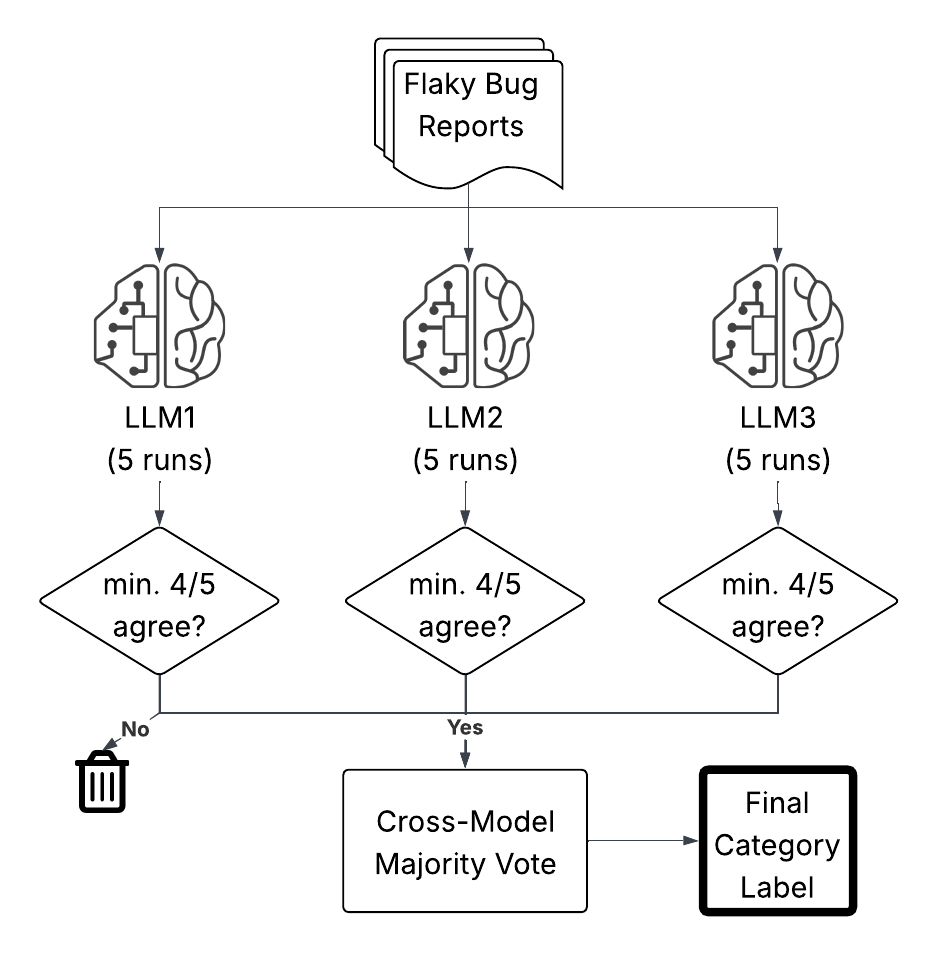}
    \caption{Our bug report labeling approach. 
    % We ask three LLMs to assign the given issue report to a flakiness category. We repeat the request for each LLM five times and consider only cases valid where the LLM provides an answer in 4 out of 5 repetitions. Based on the valid responses, we perform a majority vote between models, resulting in the final label. 
    }
    \label{fig:labeling}
\end{figure}

In addition to labeling tests as system tests or native unit tests, we combined a manual and an automated approach to categorize the root causes of the issue reports.
We manually annotated a subset of the issue reports in our dataset. For automation, we used an \emph{LLMs-as-annotators} approach as shown in \Cref{fig:labeling}. Given the description, root cause, symptom, and resolution of an issue report, we repeatedly asked three different LLMs five times to \enquote{label the report into one of the following categories} using the categories described in \Cref{sec:background}. We considered a response from an LLM \emph{valid} if it assigns the same category 4 out of 5 times. To obtain the final label, we performed a majority vote between the valid answers of the three LLMs (\texttt{gemini-2.5-pro}, \texttt{gpt-5}, \texttt{claude-4-sonnet}). We used a temperature of 0 for all three models to decrease the non-determinism of the responses~\cite{ouyang2025empirical, baltes2025guidelinesempiricalstudiessoftware}.
In an initial examination of the resulting LLM-generated labels, we found some cases where two root cause categories overlap. For example, some tests may only experience timeout issues on a specific platform. To conclude such cases, we added a code book to the prompt, which specifies the following cases: 
\begin{enumerate}
    \item Prefer \textsc{Platform} if the error occurs only in a certain setup.
    \item Prefer \textsc{Async Wait} over \textsc{Timeout} if the timeout occurs during a method call within the test.
    \item Prefer \textsc{Concurrency} over \textsc{Memory Management} if memory-related issues arise due to, e.g., race conditions.
\end{enumerate}

Finally, we quantified model-human and model-model inter-rater agreements. We first compared the labels of our three models using Fleiss' Kappa, a measure of agreement between multiple reviewers. Thus, we sought to understand the model-model agreement, which has been shown to correlate positively with human-model agreement in prior work~\cite{ahmed2025can}.
For model-model agreement, we obtained $\kappa = 0.78$, which represents very good agreement~\cite{sim2005kappa}.
All three models exhibit high internal consistency, i.e., provided the same answers in 4 out of 5 repeated requests in more than 88\% of the cases. This is expected since we set the temperature to 0 to achieve the best possible determinism in the model responses.
For the majority vote, we observed at least one valid answer for all issue reports, three valid answers for 438 of 559 reports (78\%), and two valid answers for 545 of 559 reports (97\%). In cases where only two models provided valid answers and the two answers disagreed (55), we resolved the disagreement through human judgment.
To quantify the model-human agreement in our labeling, we used Cohen's Kappa to compare the majority voting results with human annotations. We achieved $\kappa=0.63$, which represents substantial agreement~\cite{landis1977measurement}. Disagreements between the model and the human labels were mainly due to issue reports that could be assigned to multiple categories.

\subsection{RQ2: Trends and Patterns}

\subsubsection{Motivation}

\citet{parry2025systemic} coined the term \emph{systemic flakiness} to describe the co-occurrence of flaky failures, i.e., multiple tests failing flakily at the same time for the same reason. They identified network issues as a prevalent cause of such systemic flakiness and pointed out that awareness of co-occurring flakiness can reduce the cost of fixing flakiness by targeting multiple flaky tests with a single fix. With our second research question, our aim is to determine whether the root causes of flakiness differ when multiple tests are affected. In addition, we want to gain an overview of how the root causes of flaky tests change over time.

\subsubsection{Approach}

To investigate these two questions in SAP HANA issue reports, we analyzed the labeled data from RQ1. In contrast to \citet{parry2022survey}, who clustered individual tests given their likelihood of co-occurring failures, we focus on identifying systemic flakiness on a root cause category level to identify whether certain root cause categories are more likely to produce co-occurring failures. We performed two different analyses. First, we investigated the root causes of issue reports with multiple linked tests (127) to determine whether there are root cause categories with a higher chance of affecting multiple tests. Second, we aggregated the number of issue reports for each root cause category into quarterly intervals to gain insight into trends and variation in the number of issue reports across root cause categories over time. 

\begin{table}[t]
\caption{Flakiness root causes per test type in our sample of issue reports ($n=559$).} % Percentage values refer to the total value per column.
\label{tab:root-causes}
\centering
\small
\begin{tabular}{p{1.75cm} *{4}{r@{~}r}}
\toprule
\textbf{Root Cause} & \multicolumn{2}{c}{\textbf{System}} & 
\multicolumn{2}{c}{\textbf{Unit}} &
\multicolumn{2}{c}{\textbf{Mixed}} &
\multicolumn{2}{c}{\textbf{Total}} \\
\midrule
\textsc{Concurrency} & 85 & (19\%) & 42 & (39\%) & 3 & (25\%) & 130 & (23\%) \\
\textsc{Timeout} & 77 & (18\%) & 8 & (7\%) & 2 & (17\%) & 87 & (16\%) \\
\textsc{Oracle-\newline Brittleness} & 49 & (11\%) & 7 & (6\%) & 1 & (8\%) & 57 & (10\%) \\
\textsc{Configuration} & 49 & (11\%) & 6 & (5\%) & 1 & (8\%) & 55 & (10\%) \\
\textsc{Async Wait} & 42 & (10\%) & 2 & (2\%) & 1 & (8\%) & 52 & (9\%) \\
\textsc{Isolation} & 28 & (6\%) & 13 & (12\%) & 2 & (17\%) & 43 & (8\%) \\
\textsc{Platform} & 25 & (4\%) & 13 & (12\%) & 1 & (8\%) & 39 & (7\%) \\
\textsc{Application} & 35 & (8\%) & 1 & (1\%) & 0 & (0\%) & 36 & (6\%) \\
\textsc{Test-\newline Framework} & 12 & (3\%) & 5 & (5\%) & 0 & (0\%) & 17 & (3\%) \\
\textsc{Error-\newline Handling} & 14 & (3\%) & 1 & (1\%) & 1 & (8\%) & 16 & (3\%) \\
\textsc{Memory-\newline Management} & 9 & (2\%) & 4 & (4\%) & 0 & (0\%) & 13 & (2\%) \\
\textsc{Environment} & 9 & (2\%) & 2 & (2\%) & 0 & (0\%) & 11 & (2\%) \\
\textsc{Randomization} & 1 & (0\%) & 3 & (3\%) & 0 & (0\%) & 4 & (1\%) \\
\textsc{Network} & 3 & (1\%) & 0 & (0\%) & 0 & (0\%) & 3 & (0\%) \\
\textsc{Unknown} & 0 & (0\%) & 2 & (2\%) & 0 & (0\%) & 2 & (0\%) \\
\textsc{Numerical-\newline Imprecision} & 1 & (0\%) & 0 & (0\%) & 0 & (0\%) & 1 & (0\%) \\
\textsc{Locale} & 1 & (0\%) & 0 & (0\%) & 0 & (0\%) & 1 & (0\%) \\
\midrule
\textbf{Total} & 439 & (100\%) & 108 & (100\%) & 12 & (100\%) & 559 & (100\%) \\
\bottomrule
\end{tabular}
\end{table}

\section{Results}
\label{sec:results}

In this section, we present our results.

\subsection{RQ1: Root Causes}
\label{sec:rq1}
As shown in \Cref{tab:root-causes}, the most common root cause in our dataset is \textsc{concurrency}, which appears in 130 of 559 issue reports (23\%). This finding is in line with existing research that identified concurrency as a major cause of test flakiness~\cite{luo2014empirical,lam2019root,thorve2018empirical}. Comparing test types, we observe that more than one out of three issue reports for native unit tests is related to concurrency (39\%). However, for system tests, this proportion is considerably lower (19\%).
Instead, system tests appear to suffer more from timeout-related issues (18\% of issue reports) than native unit tests (7\%). This can be explained by an architectural difference in SAP HANA's test framework. The execution of native unit tests is globally canceled after one hour, with none of the native unit tests usually running for close to one hour. In contrast, the system tests were formerly assigned a dedicated per-test-timeout by developers, which was close to the test's average execution time. This architectural difference was harmonized at the beginning of 2024~\cite{berndt2024taming}, which represents approximately half of the time interval in our study. Looking at the data after harmonization, we find that the proportion of timeout-related issues for system and native unit tests is similar. 

Furthermore, while \textsc{Oracle Brittleness} and \textsc{Configuration} appear as common categories for system tests, native unit tests tend to suffer more from flakiness caused by \textsc{Isolation} and \textsc{Platform}. These findings seem intuitive. Since system tests require a running database, their result may be influenced by sophisticated configuration settings of this database instance. Based on an analysis of issue reports related to \textsc{Oracle Brittleness} of system tests, we observe that missing \texttt{ORDER BY} statements or brittle exact string comparisons are common problems.

In contrast, native unit tests are commonly used for testing low-level functionality, which can depend on specific features of the executing platform such as Non Uniform Memory Access (NUMA) support, thus explaining the high number of \emph{Platform} flakiness.

\begin{framed}
\noindent\textbf{Answer RQ1 (Root causes):} \textsc{Concurrency} is the most common root cause of flakiness for both test types in the given dataset, appearing in 130 of 559 cases (23\%). Regarding differences between test types, system tests suffer from \textsc{Timeouts} in 18\% of the reports and \textsc{Oracle Brittleness} in 11\% of issue reports. In contrast, native unit tests suffer more commonly from \textsc{Platform} (12\%) and \textsc{Isolation} (12\%).
\end{framed}

\subsection{RQ2: Trends and Patterns}
\label{sec:rq2}
With our second research question, we aim to quantify trends and patterns of root cause categories for SAP HANA's flaky tests. To achieve this, we first analyzed 127 issue reports related to more than one test. Of these 127 issue reports, 94 were related to Python tests, 21 to native unit tests, and the remaining 12 depict mixed reports linked with both Python and native unit tests. 

\begin{figure}
    \centering
    \includegraphics[width=\linewidth]{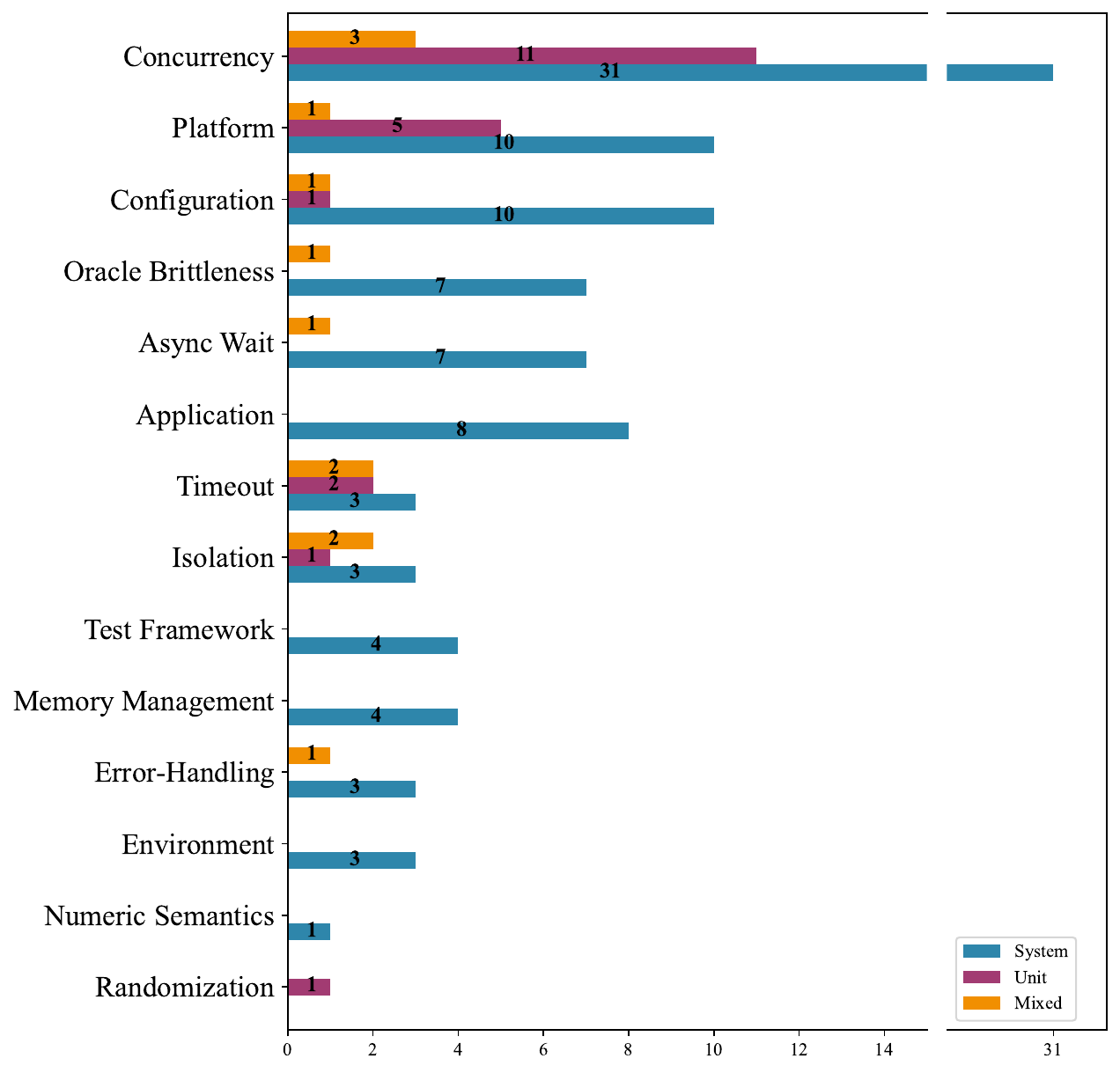}
    \caption{Barplot showing the number of issue reports per category per test type. Includes only issue reports with more than a single linked test (n=127).}
    \label{fig:multiple}
\end{figure}

\Cref{fig:multiple} shows the number of issue reports per root cause category for reports with multiple tests. \Cref{fig:multiple} shows that \textsc{Concurrency} remains the most common root cause category for both test types, being present in 45 of 127 reports (35\%). Relative to issue reports concerning only one test, we observe frequent occurrences of \emph{Platform} flakiness. In fact, the proportion of reports related to \textsc{Platform} rises from 7\% to 13\%. In contrast, the proportion of \emph{Oracle Brittleness} drops from 10\% to 6\%. The relevance of issues related to \textsc{Configuration} remains at 10\%.

\begin{figure*}
    \centering
    \includegraphics[width=\textwidth]{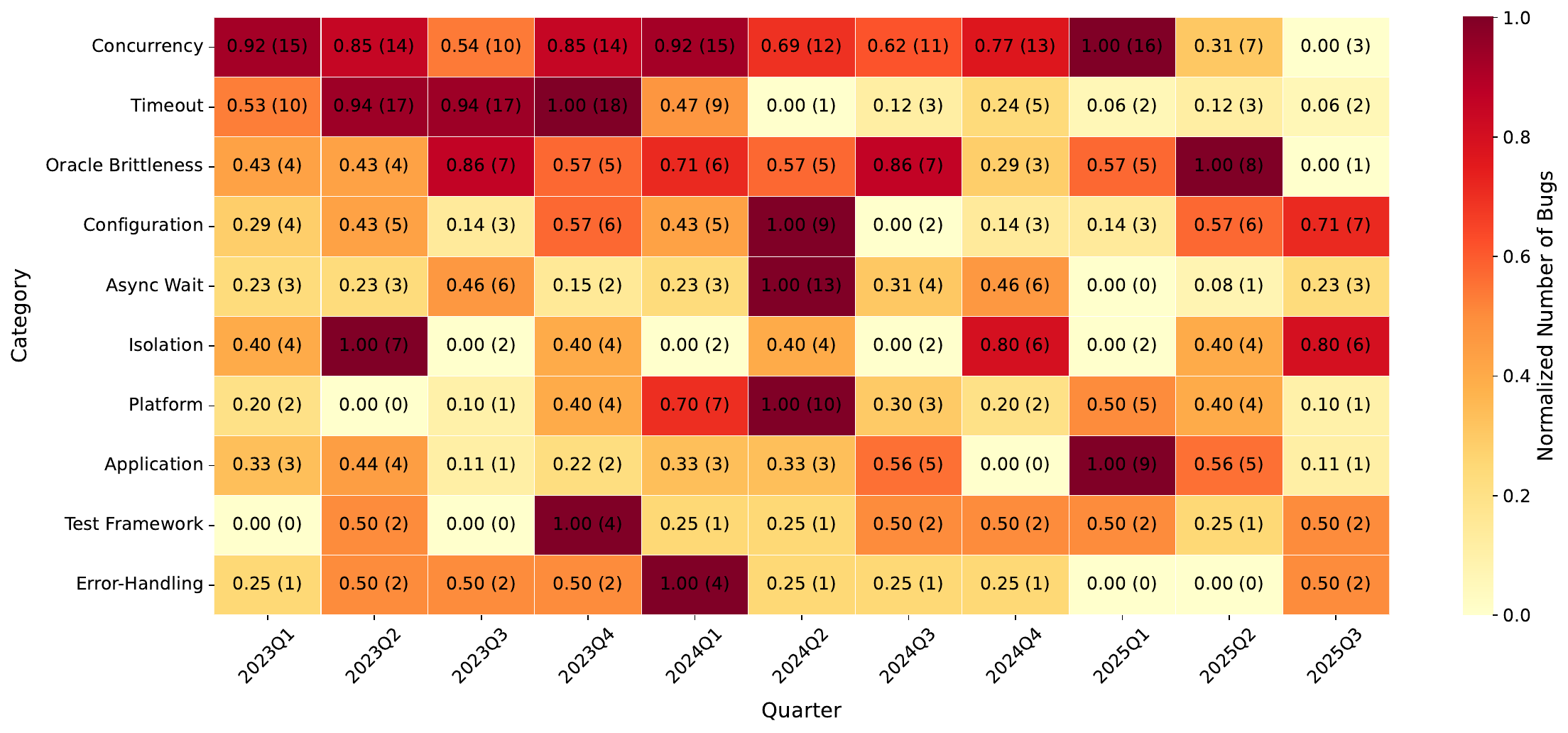}
    \caption{Heatmap showing the number of issue reports per category per quarter. Values are min-max-normalized within each category; absolute values are shown in parentheses. Note that some categories (rows) show high numbers throughout the period while others show more concentrated failure numbers.}
    \label{fig:heatmap}
\end{figure*}

\Cref{fig:heatmap} displays the number of issue reports, aggregated per category per quarter of the year. To emphasize the variations within issue reports of a category over time, we min-max-normalized the issue report counts within each category.
Overall, we observe that common root cause categories such as \textsc{Concurrency} or \textsc{Oracle Brittleness} exhibit comparatively constant rates of related issue reports throughout the examined time interval. For example, while the number of issue reports related to \textsc{Concurrency} ranges between 3 and 16 per quarter, we observe more than 10 reports per quarter in 9 out of 11 quarters. The pattern for \textsc{Oracle Brittleness} appears similar, but on a lower scale. Ranging between 1 and 8 reports per quarter throughout the examined time interval, there were 5 or more reports per quarter in 7 of 11 quarters.
Examining the number of issue reports related to \textsc{Timeout}, we observe a sharp drop after the architectural change towards a global timeout value, which confirms our expectations described in \Cref{sec:rq1}. 
Aside from \textsc{Timeout}, the most noticeable leap can be observed for \textsc{Async Wait}. Although typically ranging between 1 and 6 reports per quarter, this number rises by a factor of 2.2 for 2024-Q2 (13 issue reports). 

\begin{framed}
\noindent\textbf{Answer RQ2 (Trends):} For issue reports related to multiple tests, \textsc{Concurrency} remains the most common root cause category in the given dataset. Relative to issue reports concerning only one test, the proportion of reports related to \textsc{Platform} flakiness rose from 7\% to 12\%. Over time, the most noticeable rise in issue reports in a quarter was for \textsc{Async wait}, which rose by a factor of 2.2 compared to the other quarters. \textsc{Concurrency} and \textsc{Oracle Brittleness} showed a rather constant rate of reported issues. The rate of \textsc{Timeout} related issues dropped notably after an intervention.
\end{framed}

\section{Discussion}
\label{sec:discussion}

\emph{On LLMs as annotators.} In accordance with previous work on the use of LLMs as annotators~\cite{ahmed2025can}, we found both the resulting inter-model reliability, the human-model reliability, and the intra-model consistency helpful for signaling the reliability of the resulting labels. Overall, we consider the efficiency gains from LLM-generated labels helpful, as they enable a broad understanding of relevant root causes for flakiness at scale with reasonable effort.

\emph{On root cause labeling.} As mentioned in \Cref{sec:methodology}, we added a code book to align LLM labels with human labels, increasing the reliability of the human-model from 0.50 to 0.63. As mentioned in \Cref{sec:methodology}, the instructions in our code book addressed mainly cases where flaky failures were caused by multiple contributing factors rather than an isolated cause that could be pinpointed. For example, a test might only face timeout issues on a specific platform, or race conditions might only occur when certain environmental conditions are met. Such co-occurrences of different influencing factors can cause complex reproductions of flaky failures, which have been reported as a major problem of flakiness in previous work~\cite{eck2019understanding, lam2020understanding}. This overlap of different categories of flakiness root causes has also complicated prior work on automated flakiness categorization~\cite{rahman2025understanding}. Future work could examine the idea that root-causing flaky tests is a multi-label problem, in which flaky failures result from a combination of contributing factors. 

\emph{On the resulting root causes.} Based on the labels given by the LLMs, we identified \emph{Concurrency} as the most common root cause in the given dataset. This result seems intuitive, since compared to other software projects, SAP HANA is highly dependent on parallelization within the product code as well as for testing. Regarding the generalizability of our results, it is important to note that the ticket system from which our issue reports are drawn is primarily used by developers. As a result, global issues related to the test infrastructure are likely underrepresented in our dataset, as problems that developers can address are typically reported.

\section{Conclusions}
\label{sec:conclusion}
In this study, we empirically analyzed issue reports related to flakiness in the context of SAP HANA. To gain an overview of prevalent root causes of flakiness in that context, we explored an \emph{LLMs as annotators} approach to divide issue reports into different root cause categories derived from previous work. Based on a comparison with a manually labeled sample, our results suggest that LLMs exhibit reasonable human-model alignment when automatically categorizing issue reports related to flakiness. 

As previous research has suggested that the root causes of flakiness are context-dependent, we analyzed the root causes for unit and system tests separately. Our analysis reveals that both test types are primarily affected by \textsc{Concurrency} issues, including race conditions and improper thread handling. Regarding the differences between the tests, our results indicate that SAP HANA's system tests tend to experience flaky failures due to \textsc{Oracle Brittleness}, caused by improper use of \texttt{ORDER BY} or brittle exact string matching. In contrast, unit tests are more susceptible to \textsc{Platform}-specific details, such as compiler differences.  

We encourage future work to elaborate on the idea of flakiness root causes as a multi-label problem. Instead of pinpointing flakiness down to a single root cause, it is valuable to gain more insights into interdependencies that cause flakiness. We believe that, especially in the context of large-scale software projects, identifying common problems caused by such interdependencies can help build future tooling to detect and fix flaky tests or flaky test executions.

\bibliographystyle{ACM-Reference-Format}
\bibliography{bibliography}

\end{document}